\documentclass[doublecol]{epl2}
\title{Nonextensive triplet in geological faults system}
\shorttitle{Nonextensive triplet in geological faults system} %Insert here a short version of the title if it exceeds 70 characters

\author{D. B. de Freitas\inst{1} \and G. S. Fran\c{c}a\inst{2} \and T. M. Scheerer\inst{2} \and C. S. Vilar\inst{3} \and R. Silva\inst{1,}\inst{4}}
\shortauthor{D. B. de Freitas \etal}

\institute{
  \inst{1} Departamento de F\'{\i}sica,
    Universidade Federal do Rio
    Grande do Norte, 59072-970
    Natal,  RN, Brazil\\
  \inst{2} Observatorio Sismologico-IG/UnB, Campus Universit\'ario Darcy Ribeiro SG 13 Asa Norte, 70910-900 Bras\'{\i}lia, Brazil \\
  \inst{3} Instituto de F\'{\i}sica, Universidade Federal da Bahia, Campus Universit\'ario de Ondina, 40210-340 Salvador, Brasil\\
  \inst{4} Universidade do Estado do Rio Grande do Norte, UERN,
Departamento de F\'{\i}sica, Mossor\'o -- RN, CEP 59610--210, Brazil
}

\abstract{
The San Andreas fault (SAF) in the USA is one of the most investigated self-organizing systems in nature. In this paper, we studied some geophysical properties of the SAF system in order to analyze the behavior of earthquakes in the context of  Tsallis's $q$--Triplet. To that end, we considered 134,573 earthquake events  in  magnitude interval $2\leq m<8$, taken from the Southern Earthquake Data Center (SCEDC, 1932 - 2012). The values obtained (``$q$--Triplet''$\equiv$$\{$$q$$_{stat}$,$q$$_{sen}$,$q$$_{rel}$$\}$) reveal that the $q_{stat}$--Gaussian behavior of the aforementioned data exhibit long-range temporal correlations. Moreover, $q_{sen}$ exhibits quasi-monofractal behavior with a Hurst exponent of 0.87.}

\begin{document}

\maketitle

\section{Introduction}
Earthquakes are among the most complex spatiotemporal phenomena investigated in the context of self-organized criticality (SOC), introduced in Ref. \cite{bak87}. In this regard, let us consider the so-called fault systems, a complex phenomenon related to the deformation and sudden rupture of some parts of the Earth's crust driven by convective motion in the mantle. One of the first examples of self-organizing systems in nature \cite{rundle} is the San Andreas Fault (SAF) in California. The SAF,  one of the world's longest and most active geological faults,  is  $\sim$1200 Km long, $\sim$15 Km deep, and about  20 million years old. It forms the boundary between the North American and Pacific plates and is classified as a right lateral strike-slip fault, although its movement also involves comparable amounts of reverse slip \cite{12}. From the geophysical standpoint , a considerable number of investigations have been conducted in order to better understand the complexity of this system (see, e.g., \cite{schulz1997} and references therein). In contrast to the complexity of earthquakes,  empirical laws are extremely simple, e.g. the Gutenberg-Richter law, which gives the number of earthquakes with a magnitude $M>m$ \cite{GR44}, and the Omori law for temporal distribution of aftershocks \cite{omori}.

Several studies have demonstrated  that seismicity exhibits an out-of-equilibrium behavior that is  being investigated by different authors, e.g. studies based on  wavelet-based multifractal analysis \cite{enescu2004} and  nonextensive statistical mechanics \cite{costa04,silva06,franca2007}, among others. In the present study, we consider nonextensive formalism, which is a generalization of Boltzmann-Gibbs statistical mechanics (B-G statistics) for out-of-thermal equilibrium systems and described by the \textit{entropic parameter} $q$. The celebrated Boltzmann-Gibbs (B-G) statistics is recovered at $q=1$ \cite{tsallis1988,abe01,gell2004}. This parameter measures the degree of nonextensivity in the stochastic process.

Tsallis statistics is based on the $q$-exponential and $q$-logarithm, two central functions defined by

\begin{equation}
\label{tsallisa}
\exp_{q}(f)=[1+(1-q)f]^{1/1-q},
\end{equation}
and
\begin{equation}
\label{tsallisb}
\ln_{q}(f)=\frac{f^{1-q}-1}{1-q},
\end{equation}
which produces entropy $S_{q}$ \cite{gell2004}, associated with $q$-statistics,

\begin{equation}
\label{tsallis1}
S_{q}=k\frac{1-\int [PDF(x)]^{q}\mathrm dx}{q-1} \quad\ (q\in \texttt{R}),
\end{equation}
where the Boltzmann-Gibbs entropy, usual exponential and logarithm are recovered if $q=1$.

\begin{figure}
\resizebox{0.45\textwidth}{!}{%
  \includegraphics{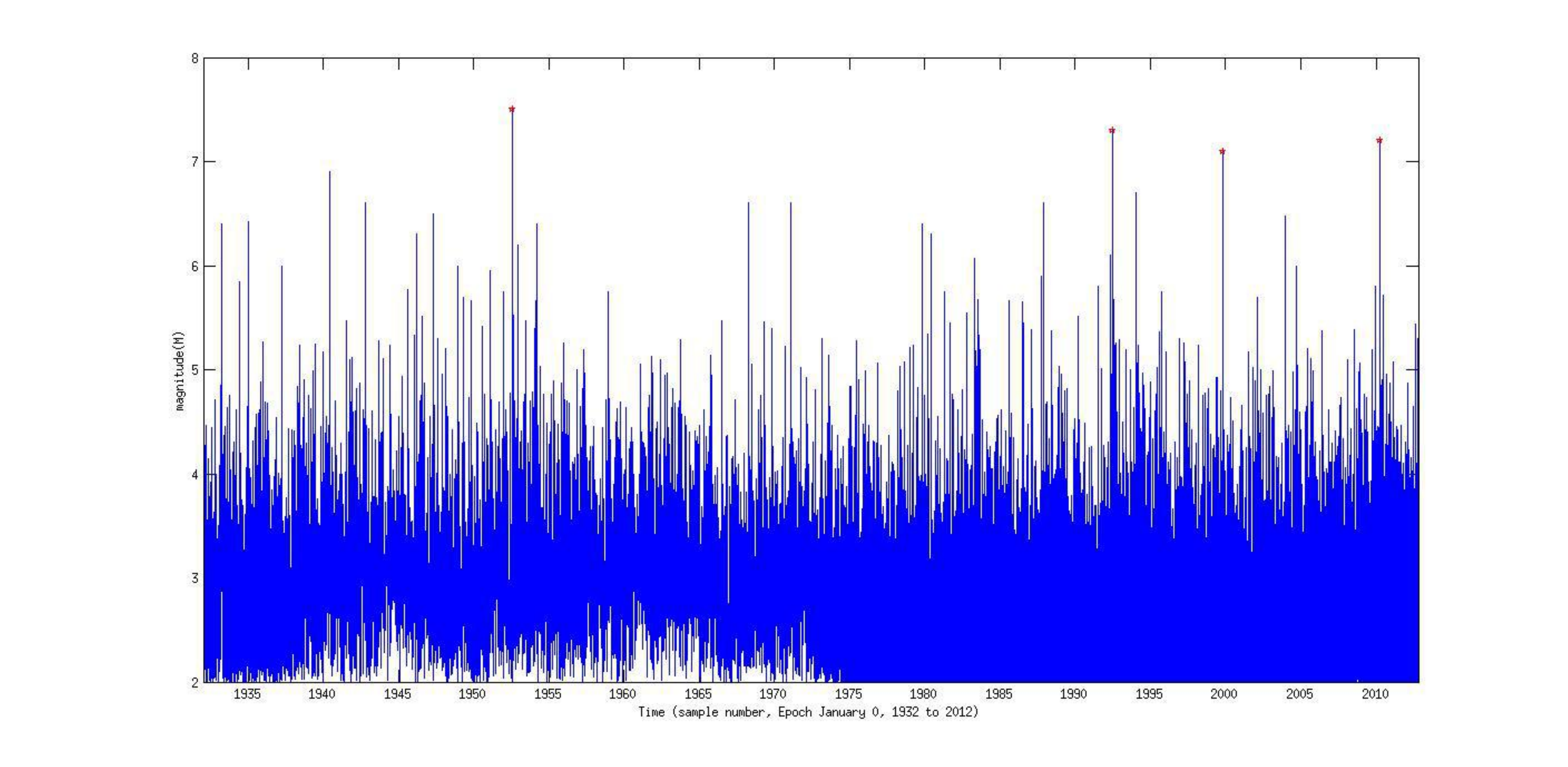}
  }
%\vspace{5cm}       % Give the correct figure height in cm
\caption{Time series for magnitude of earthquakes along SAF. The peaks denote the maximum magnitudes.}
\label{fig1}       % Give a unique label
\end{figure}

\begin{figure}
\resizebox{0.49\textwidth}{!}{%
  \includegraphics{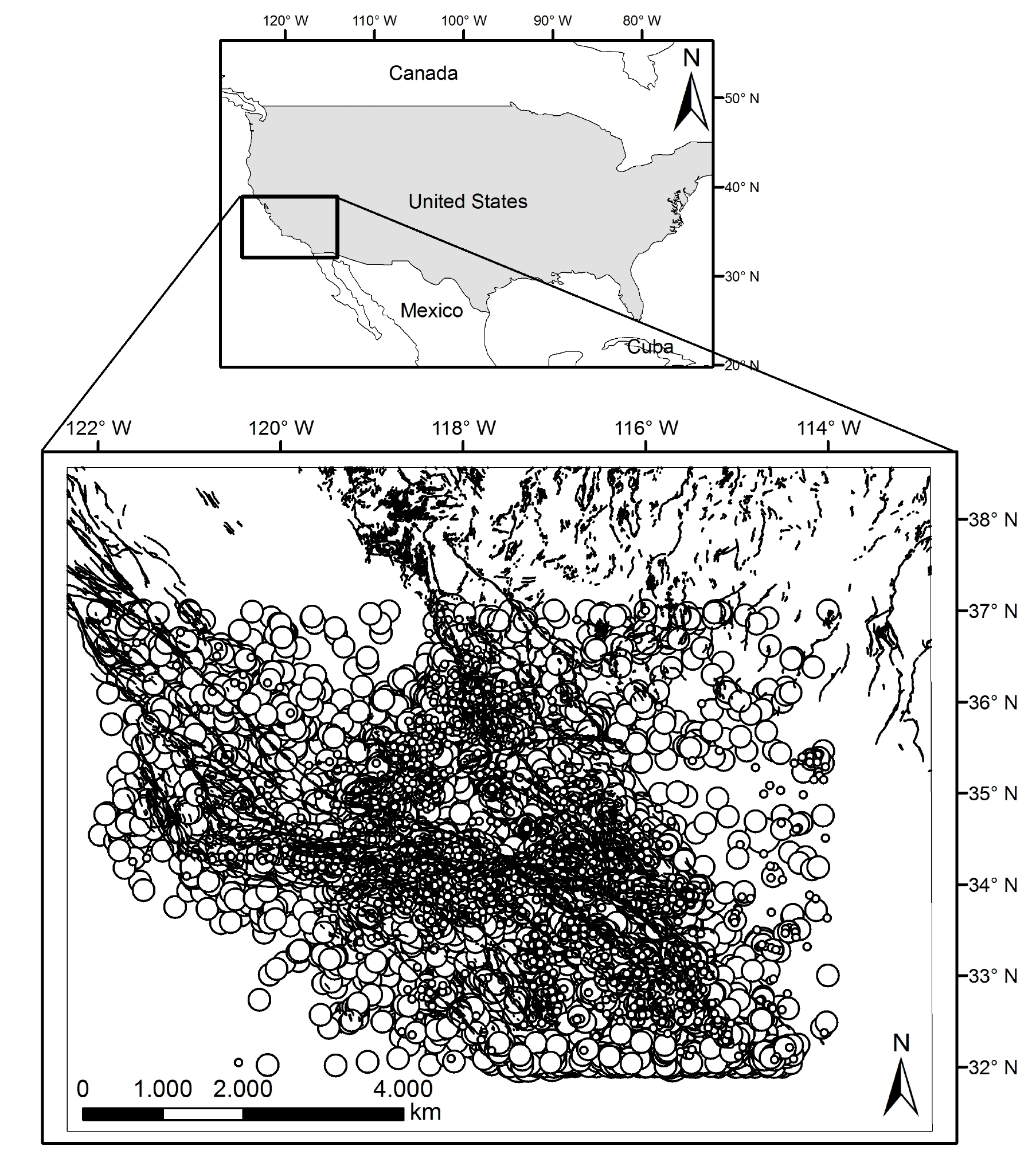}
  }
%\vspace{5cm}       % Give the correct figure height in cm
\caption{Map of seismicity in SAF system, showing epicenters of earthquakes considered in this study (source SCEDC).}
\label{fig2a}       % Give a unique label
\end{figure}

\begin{figure}
\resizebox{0.49\textwidth}{!}{%
  \includegraphics{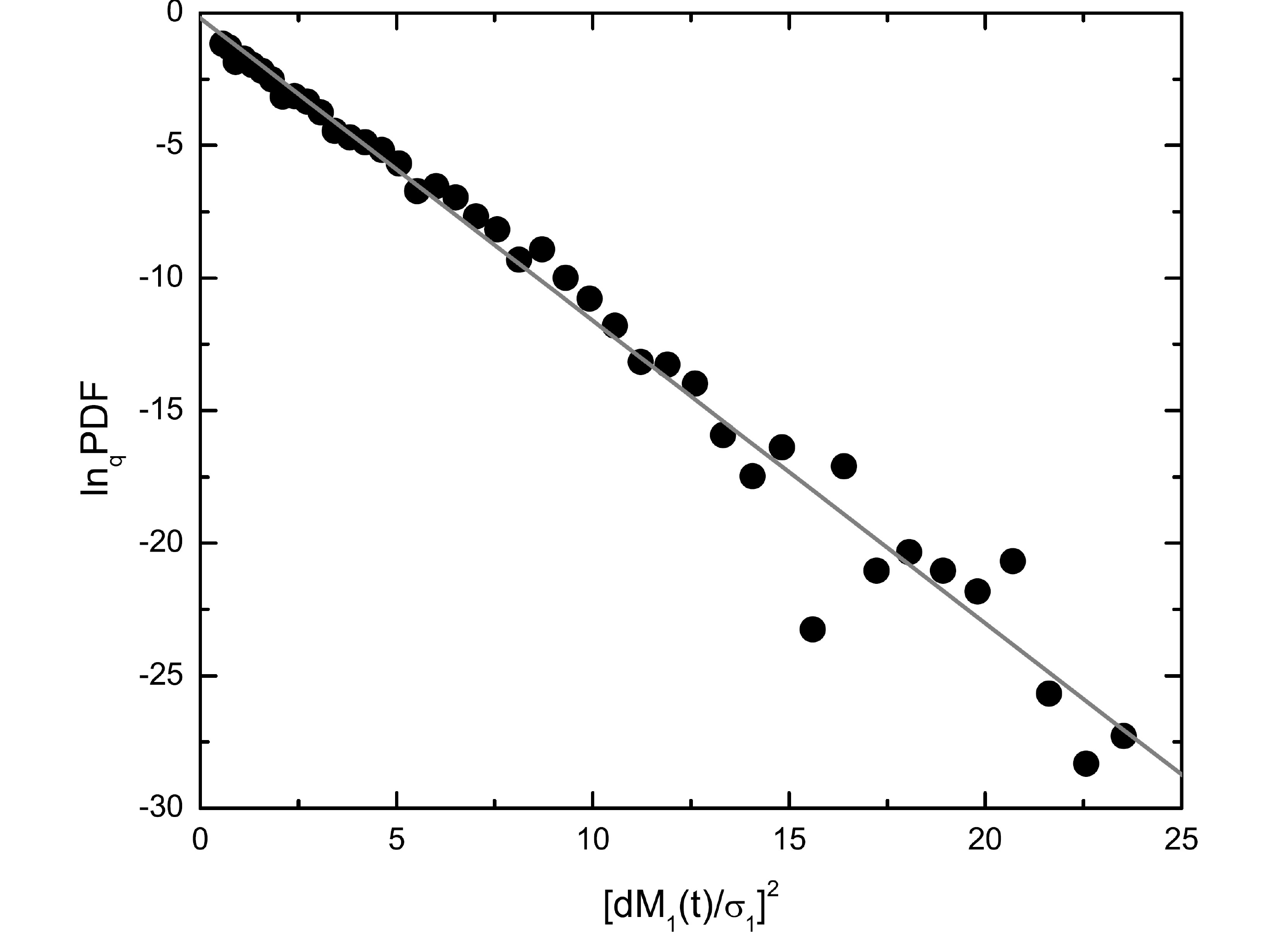}
  }
%\vspace{5cm}       % Give the correct figure height in cm
\caption{Linear correlation between $\ln_{q}[PDF]$ and $[\mathrm dM_{1}(t)/\sigma_{1}]^{2}$, where $q_{stat}=1.364\pm0.04$, with $R^{2}=0.992$ and $\chi^{2}/dof=7.0236\times10^{-6}$.}
\label{fig2}       % Give a unique label
\end{figure}

\begin{figure}
\resizebox{0.49\textwidth}{!}{%
  \includegraphics{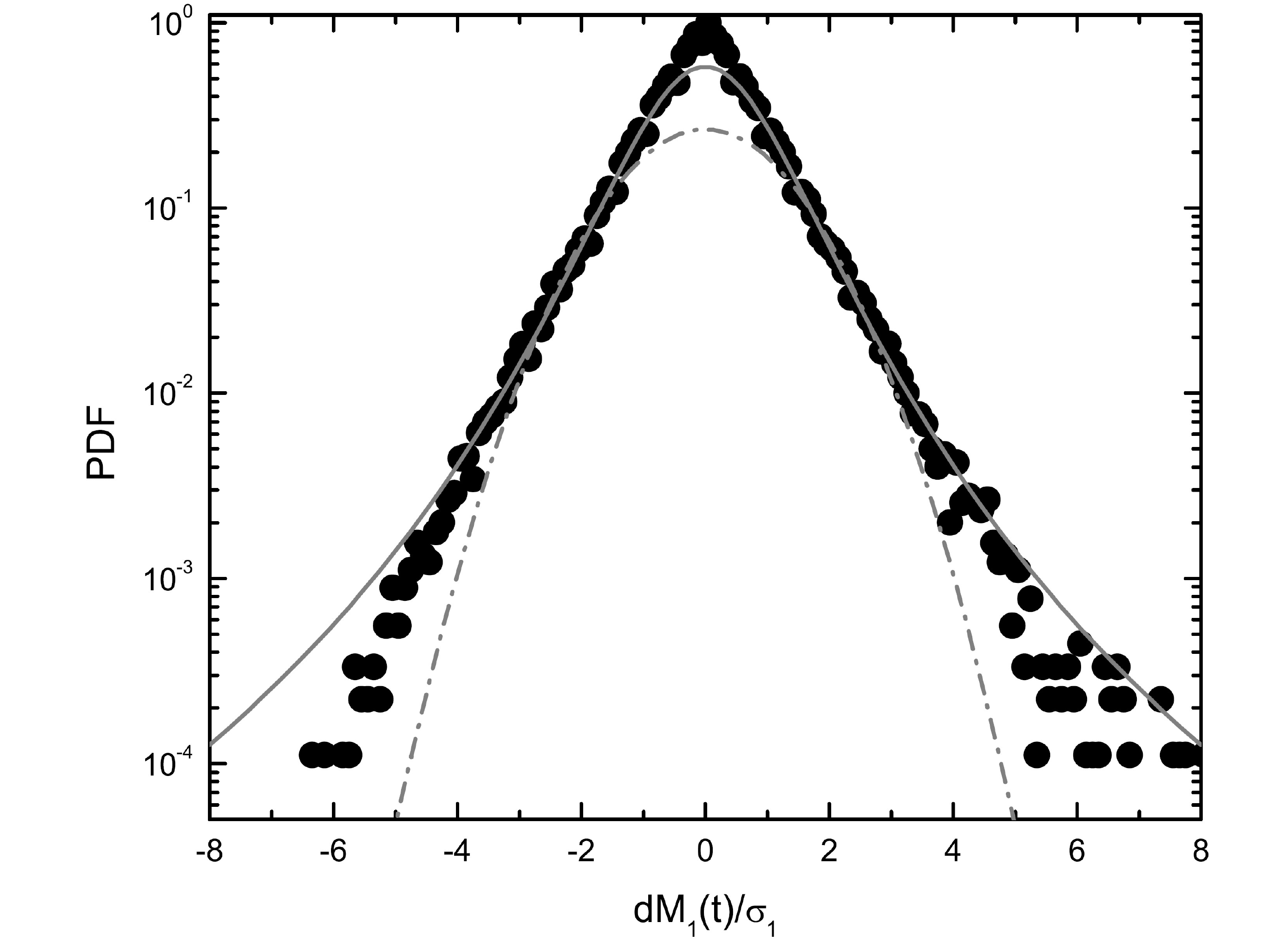}
  }
%\vspace{5cm}       % Give the correct figure height in cm
\caption{In black circles is ploted distribution of increment for SAF data. In solid black line, the $q_{stat}$-Gaussian distribution based in eq. (\ref{1pdf}) with $B_{q}=0.858\pm0.16$. In dashed line, the best fit with a standard Gaussian.}
\label{fig3}       % Give a unique label
\end{figure}

\begin{figure}
\resizebox{0.49\textwidth}{!}{%
  \includegraphics{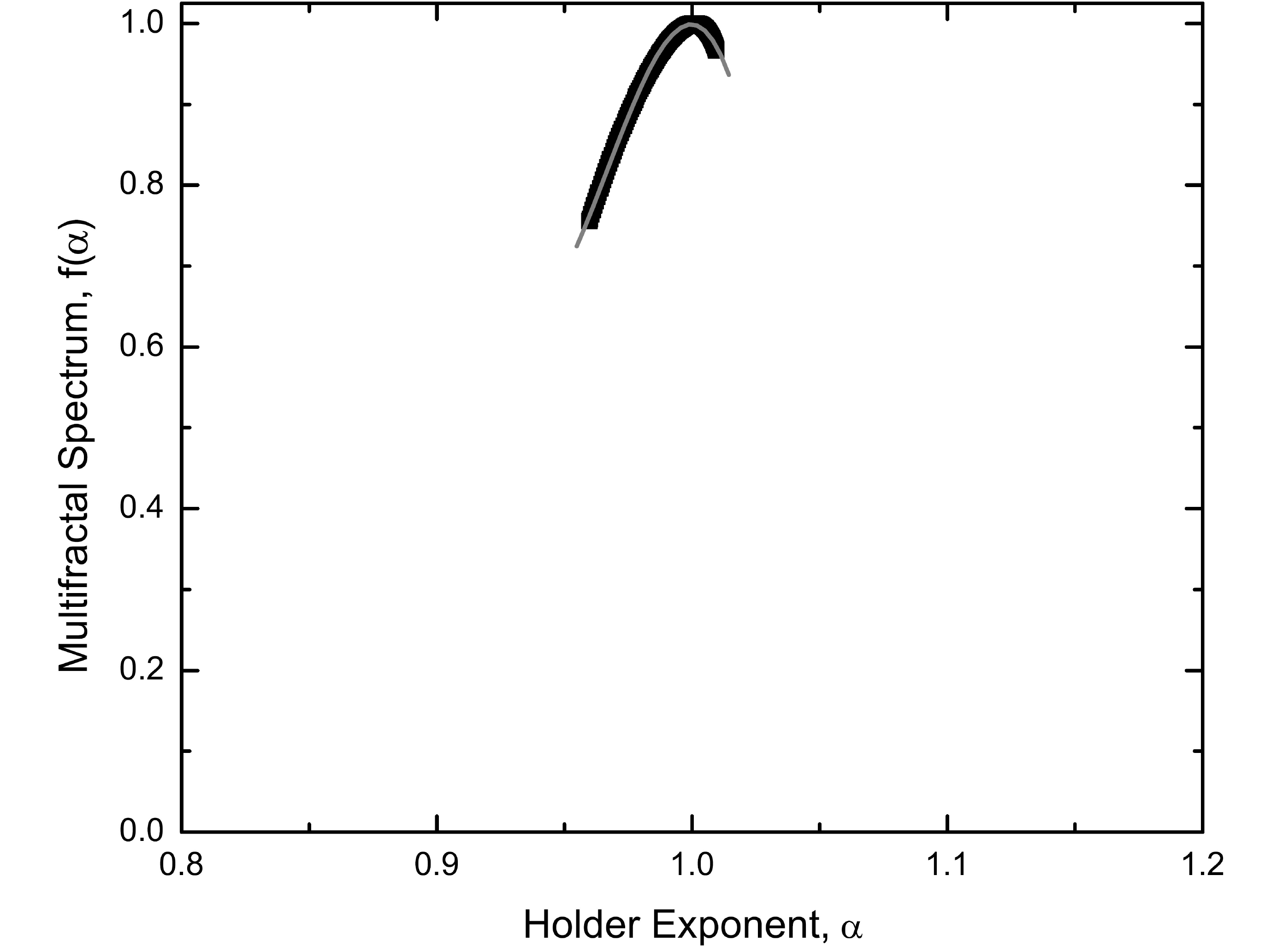}
  }
%\vspace{5cm}       % Give the correct figure height in cm
\caption{The symbols are based on measurements of multifractal spectrum $f(\alpha)$ $vs.$ $\alpha$ obtain from $M(t)$. We obtain for SAF $\alpha_{min}=0.924\pm 0.04$ and $\alpha_{max}=1.051\pm 0.11$ a $q_{sen}=-6.747\pm 0.35$. The curve represent the best ajustment using a cubic fit to the data.}
\label{fig4}       % Give a unique label
\end{figure}

\begin{figure}
\resizebox{0.49\textwidth}{!}{%
  \includegraphics{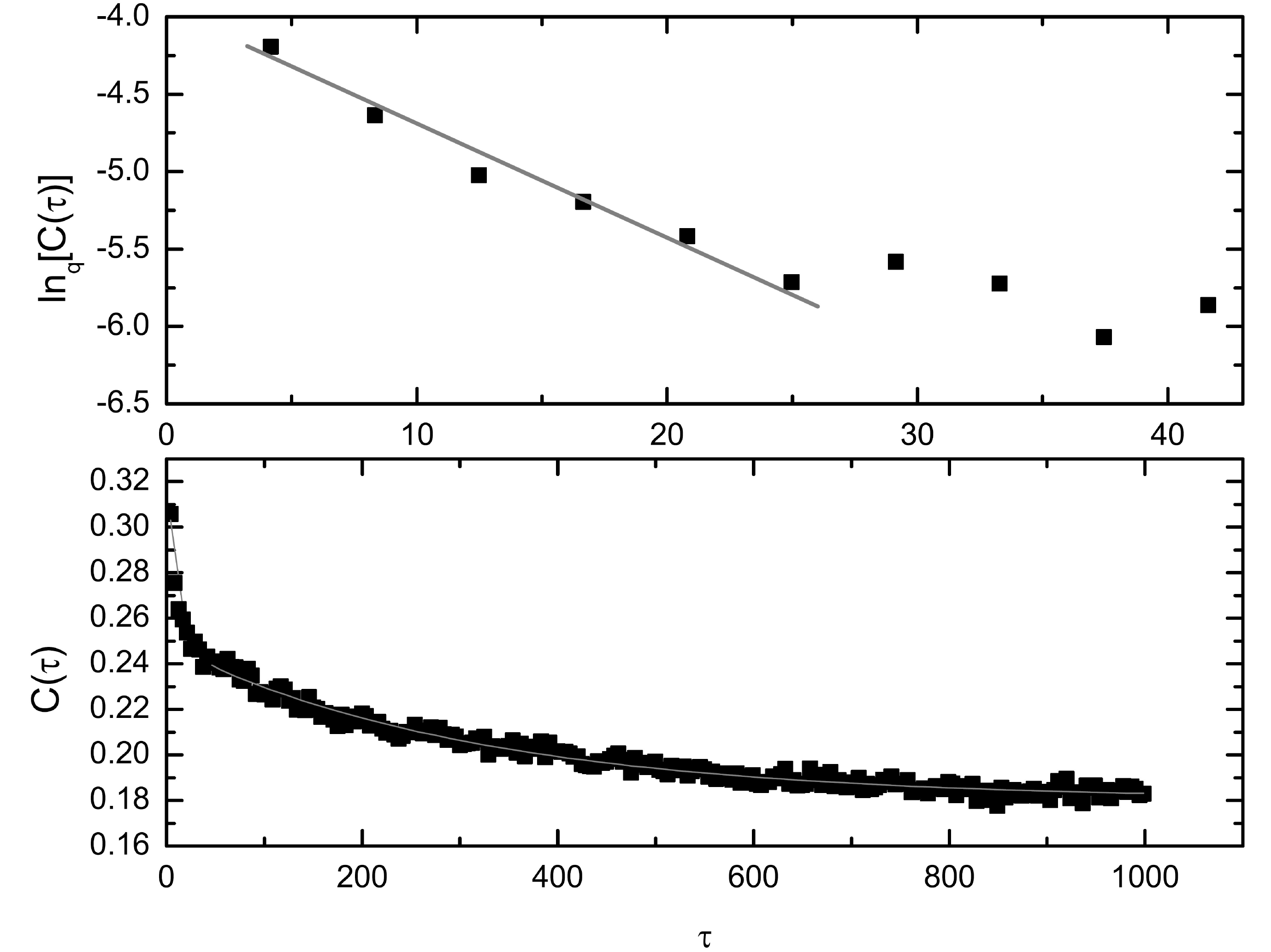}
  }
%\vspace{5cm}       % Give the correct figure height in cm
\caption{\textit{Upper panel}: $\ln_{q}$ of autocorrelation coefficient $C(\tau)$ \textit{vs.} time delay $\tau$ for SAF data. \textit{Lower panel}: symbols represent the autocorrelation function for our sample and the gray line represents a double exponential fit with characteristic times of $t_{1}=8.42$ and $t_{2}=313.74$ yielding a ratio about 37 between these two time scales ($R^{2}=0.964$, $\chi^{2}/dof=1.4\times 10^{-3}$ and time is expressed in order of hours).}
\label{fig5}       % Give a unique label
\end{figure}

This theory has been successfully applied to many complex physical systems such as geological faults \cite{franca2007} and astrophysical systems \cite{burlaga2004,burlaga2009,defreitas2009}. In 2004, Tsallis \cite{tsallis2004} proposed the existence of a three-parameter set ({\it q}$_{stat}$,{\it q}$_{sen}$,{\it q}$_{rel}$), also known as $q$-Triplet, characterized by metastable states in nonequilibrium, where $q_{stat}>1$, $q_{sen}<1$ and $q_{rel}>1$. When ({\it q}$_{stat}$,{\it q}$_{sen}$,{\it q}$_{rel}$)=(1,1,1), the set denotes the B--G thermal equilibrium state. Burlaga and Vi{\~n}as \cite{burlaga2005}  used this triplet to describe the behavior of two sets of daily magnetic field strength performed by Voyager 1 in the solar wind in 1989 and 2002. In 2009, de Freitas and De Medeiros \cite{defreitas2009} presented a physical corroboration of the $q$--Triplet, based on analyses of the behavior of three sets of daily magnetic field strength observed by different solar indices. More recently, Ferri, Savio and Plastino \cite{ferri2010} showed a physical implication of this triplet for the ozone layer in Buenos Aires, Argentina.

The main aim of this study is to analyze the behavior of physical parameters directly reflecting seismic activity in the context of Tsallis $q$--Triplet's formalism, and to compare the properties of this $q$--Triplet with those  expected for a metastable or quasi-stationary dynamical system described by nonextensive statistics. In this context, we focus our attention on the magnitude-values for SAF data $M(t)$ and their  hourly variability $\mathrm dM_{\tau}(t)$. Following the ideas presented in Ref. \cite{caruso2007}, we focus our investigation on the ``return'' or fluctuation $\mathrm dM_{\tau}(t)=M(t+\tau)-M(t)$, which denotes the differences between ``avalanche'' sizes obtained at time $t+\tau$ and at time $t$. With respect to seismic activity, this analysis also checks the validity of $q$--Central Limit Theorem, the so-called $q$--CLT, recently conjectured by Umarov, Tsallis and Gell--Mann \cite{umarov2008}.

The remainder of this paper is organized as follows: in Section 2, we present our seismic sample; the main results and discussions are presented in section 3; and, finally, conclusions are put forth in the last section.

%Estou aqui

\section{The seismic data}
Fig. \ref{fig1} shows the time series for magnitude $M$ of earthquakes along the SAF,  in the interval $2\leq M<8$, with 134,573 events. These were taken from the Southern California Earthquake Data Center (SCEDC) for 1932 to 2012. This range was chosen because for small magnitudes it has the limitation of seismic monitoring in the area, since many such events are unregistered. The lower panel of this figure shows non-overlapping magnitude fluctuations (return) in $M$ for $\mathrm d M_{1}(t)$. Fig. \ref{fig2a} illustrates the distribution of events considering the SAF map .

Fig. \ref{fig2a} shows the data and the San Andreas fault system. This system is more than 800 miles long and extends to depths of at least 10 miles. The fault is a complex zone of crushed and broken rock ranging from a few hundred feet to a mile wide. Many smaller faults branch from and join the San Andreas fault zone. Almost any road cut in the zone shows a myriad of small fractures, fault gouge (pulverized rock), and a few solid pieces of rock \cite{schulz1997}. The movement that occurs along the fault is a right-lateral strike-slip  forming the tectonic boundary between the Pacific Plate and the North American Plate. 

\section{Results and discussions}
In this section, we show results after the estimation of ``$q$--Triplet''$\equiv$$\{$$q$$_{stat}$,$q$$_{sen}$,$q$$_{rel}$$\}$ based on SAF data from 1932 to 2012 (see Fig. \ref{fig1}). These results are presented in three subsections, each associated to the properties of one of the $q$'s.

\subsection{On the behavior of the $q$-stationary parameter}
For time series $M(t)$, increment fluctuations due to its variability over timescale $\tau$ is given as $\mathrm dM_{\tau}(t)=M(t+\tau)-M(t)$. The values of $q_{stat}$ are derived from probability distribution functions (PDFs). These PDFs are obtained from the variational problem using the continuous version for the nonextensive entropy given by eq. (\ref{tsallis1})
\begin{equation}
\label{1pdf}
PDF=A_{q}\left[1+(q-1)B_{q}\mathrm dM_{\tau}(t)^{2}\right]^{\frac{1}{1-q}},
\end{equation}
the {\it entropic parameter} $q$ is related to the size of the tail in the distributions \cite{burlaga2009} and coefficients $A_{q}$ and $B_{q}$ for $q>1$ are given by
\begin{equation}
\label{2pdf}
A_{q}=\frac{\Gamma\left[\frac{1}{q-1}\right]}{\Gamma\left[\frac{3-q}{2q-2}\right]}\sqrt{\frac{q-1}{\pi}B_{q}}
\end{equation}
and
\begin{equation}
\label{3pdf}
B_{q}=\frac{1}{\left[(3-q)\sigma^{2}_{q}\right]},
\end{equation}
for further details see Ref. \cite{queiros2007}.

Following the same procedure described by \cite{ferri2010}, we varied the index $q$ between 1.0 and 2.0, making a linear adjustment in each computational iteration  and evaluating the specific correlation coefficient $R^{2}$. The best linear fit is obtained for $q_{stat}=1.364\pm0.04$ with $R^{2}=0.992$ as shown in Fig. \ref{fig2}. It should be emphasized that this $q_{stat}$ value is fully consistent with the bounds obtained from several independent studies involving the  nonextensive Tsallis framework (see, e.g. \cite{19}). The PDF for the return $\mathrm d_{\tau}M(t)$ on scale $\tau=1$ is shown in Fig. \ref{fig3}. On this scale we can conduct a closer investigate of a possible correlation between events . Our study used the Levenberg--Marquardt method \cite{leven1944,marq1963} to compute  PDFs with symmetric Tsallis distribution from Equation (\ref{1pdf}). In this adjustment, we found $B_{q}=0.858\pm0.16$. These results are consistent with the value expected for nonlinear systems, where the random variable is the sum of strongly correlated contributions \cite{burlaga2009,burlaga2005,tirnakli2007}. In this respect, we showed that PDFs for the return $\mathrm dM_{1}(t)$ have fat tails with a $q$-Gaussian shape.

\subsection{On the behavior of the $q$-sensibility parameter}
Values of the $q_{sen}$-index are directly related to system instability and entropy growth. These values can be obtained from  multifractal (or singularity) spectrum $f(\alpha)$, where $\alpha$ is the singularity strength or H\"{o}lder exponent. Spectrum $f(\alpha)$ is derived via a modified Legendre transform, through the application of the MF-DFA5 method \cite{kantel2002}. This method consists of a multifractal characterization of a nonstationary time series, based on a generalization of detrended fluctuation analysis (DFA). MFDFA performs best when the signal is a noise-like time series. However, there is also  difficulty in visualizing the difference between walk and noise-like time series. As suggested by \cite{eke2002}, before application, it is necessary to run a DFA and  verify if the value of the Hurst exponent is less than 1.2. For  SAF data we obtain a Hurst exponent of 0.87, indicating that the MFDFA method can be employed directly without transformation of the time series.

The $q_{sen}$-index denotes sensitivity at initial conditions. For present purposes, we used the expression defined by Lyra and Tsallis \cite{lyra1998} for the relation between $q_{sen}$ and multifractality in dissipative systems, as follows:
\begin{equation}
\label{1sen}
\frac{1}{1-q_{sen}}=\frac{1}{\alpha_{min}}-\frac{1}{\alpha_{max}},
\end{equation}
where $\alpha_{min}$ and $\alpha_{max}$ denotes the roots of the
best-fit.

The multifractal characterization of these data is shown in Figure \ref{fig4}. These spectra $f(\alpha)$, calculated for SFA data, show a narrow H\"{o}lder exponent interval with $\alpha_{mim}$=0.924$\pm$0.04 and $\alpha_{max}$=1.051$\pm$0.11. For  multifractal spectrum width, we obtained $\Delta\alpha$=$\alpha_{max}-\alpha_{min}$, resulting in a value of 0.127. Using Equation \ref{1sen}, we found that $q_{sen}=-6.647\pm0.35$. This negative value indicates that its distribution exhibits weak chaos \cite{tsallis2004} in the full dynamical space of the system \cite{tsallis2004,burlaga2005}. Furthermore, this figure reveals that the behavior of our sample is similar to that of a monofractal-like time series.

\subsection{On the behavior of the $q$-relaxation parameter}
The value of $q_{rel}$, which describes a relaxation process, can be computed from an autocorrelation coefficient as a function of scale $\tau$ defined by
\begin{equation}
\label{1rel}
C(\tau)=\frac{\left\langle [S(t_{i}+\tau)-\left\langle
S(t_{i})\right\rangle][S(t_{i})-\left\langle
S(t_{i})\right\rangle]\right\rangle}{\left\langle [S(t_{i})-\left\langle
S(t_{i})\right\rangle]^{2}\right\rangle}.
\end{equation}

In agreement with  Tsallis statistics, we can estimate the value of $q_{rel}$ by best fit on $\ln_{q}C(\tau)$ $vs.$ scale $\tau$, as shown in Fig. \ref{fig5} (upper panel), where $C(\tau)$ is given by Equation (\ref{1rel}). In the nonextensive theory, this coefficient should decay following a power law, with increasing $\tau$, where slope $s$ is given by $s=1/(1-q_{rel})$. From this adjustment, we obtain $q_{rel}=2.69\pm0.13$ for SAF data. Moyano \cite{moyano2006} suggests that the above procedure for calculating $q_{rel}$ only be used to describe stochastic processes with linear correlations. In other words, autocorrelation coefficient $C(\tau)$ is not a good alternative to conveniently describe the non-linearity of a sample \cite{defreitas2009}.

On the other hand, in B--G statistics, in contrast to the nonextensive theory,  coefficient $C(\tau)$ should decrease exponentially with an increasing $\tau$, following a $C(\tau)=A_{1}\exp(-\tau/t_{1})+A_{2}\exp(-\tau/t_{2})$ relation, with $t_{1}$ and $t_{2}$ corresponding to the correlation or relaxation times. The fit shown in Fig. \ref{fig5} (lower panel) reveals that $t_{2}\gg t_{1}$. As mentioned by \cite{queiros2007}, this behavior is related to local equilibrium, and then a much slower decay for larger $\tau$. In agreement with these authors, this constitutes a necessary condition for the application of the superstatistical model, as described in Ref. \cite{beck2003}.

See \cite{pavlos2012} for further details and an extensive discussion about the estimation of Tsallis $q$-triplet.

\section{Conclusions}
We used a new approach to nonextensive formalism for hourly measurements of earthquakes along the SAF from 1932 to 2012. From these data we were able to estimate the values of the nonextensive three-index. We found that $q_{stat}=1.364\pm0.04$, $q_{sen}=-6.647\pm0.35$ and $q_{rel}=2.69\pm0.13$. It is important to underscore that the result of the $q_{stat}$ is consistent with the upper limit $q<2$ obtained from several independent investigations \cite{19}. In addition, the values of this triplet confirm the general scheme  $q_{sen}\leq 1 \leq q_{stat}\leq q_{rel}$, according to the nonextensive scenario proposed by Tsallis \cite{tsallis2004}. These results reveal that this system is consistent with a nonequilibrium state, strongly suggesting that long-range correlations exist among the random variables involved in the physical process that controls seismic activity.

Finally, it is worth mentioning that the nonextensive three-index can be recalculated by considering a spatiotemporal analysis for earthquakes along the SAF. This issue will be addressed in a forthcoming communication.

\acknowledgments
Research activity at the Stellar Board of the Federal University of Rio Grande do Norte (UFRN) and Federal Institute of Rio Grande do Norte (IFRN) are supported by continuous grants from CNPq and FAPERN Brazilian agency. The authors would like to thank Cesar Garcia Pav\~ao for their many
helpful with maps this work.

\end{document}